# Model Assisted Data Integration: An unbiased sampling strategy to use nonprobability data

Martin Hyllienmark, Statistics Sweden, martin.hyllienmark@scb.se

Gustaf Strandell, Statistics Sweden, gustaf.strandell@scb.se

**Abstract**
The aim of survey statistics is to produce estimates with a minimal bias and a corresponding acceptable variance given a specific budget, preferable with a minor response burden for the participants. In recent years, considerable efforts have been made to achieve this through the extended use of found or non-probability data. However, to be able to safely utilize such data, rigorous theoretical foundations is needed, where one main concern is the of lack control due to not having access to the selection mechanism for the data. Several methods have been proposed in the literature to deal with this, though often relying on assumptions that may be difficult or impossible to verify in practice. Extending on the Data Integrated (DI) estimator introduced by Kim and Tam (2021), this paper introduce the Model Assisted Data Integration (MADI) sampling strategy. The proposed sampling strategy includes an estimator that has the desired properties: it is design-unbiased, has a design-unbiased variance estimator and is suitable for the intense production cycle of the statistical agency. The estimator uses nonprobability data combined with a probability sample that has a sampling design which aims to include individuals not captured by the nonprobability data. The estimator can use arbitrary machine learning models to produce unbiased estimates. A main conclusion of the paper is that the proposed sampling strategy can produce estimates with much lower variances compared to traditional survey estimators, and we use real empirical data to illustrate this point.

## 1. Introduction

Since the end of World War II, during which electronic computers were first introduced in submarines to solve the trigonometric problem of firing torpedoes at moving targets, the use of digital technologies for civil applications has skyrocketed. Within the field of statistics, this progress has among other things provided electronic databases, effective environments for computations, web-based questionnaires and online dissemination of statistics.

Over the last 30 years, one main channel in the flooding digitalization of the civil world has roared with the integration of digital networks for all manners of interactions between objects commonly under consideration in statistics: individuals, companies, organizations, machines and even animals, resulting in abundance of new data. Cash register data, credit card data, digital accounting reports, traffic sensor data and job vacancy data are some examples of data sources that have been or are being considered for official statistics. For the statistical offices, the hope is that the use of these new sources will bring lower costs, higher quality and a reduced response burden (Tam & Holmberg, 2020).

In this paper, we define a probability sample as a sample drawn from a frame where each unit has a known positive probability of being selected to the sample by some random selection mechanism. The frame from where we draw units does not necessarily need to coincide with the population which we aim to produce statistics about. Furthermore, we will use the term nonprobability data (NPD) to refer, somewhat loosely, any data source that does not have known non-zero inclusion probabilities or is a register with complete coverage of the target population. There is a growing necessity for methods and survey designs in which data from different sources can be combined, including traditional surveys, statistical registers and non-probability data sources (Holmberg, 2025). Accordingly, considerable research activity has been directed towards multi-source statistics and data integration. There are several excellent introductions and general overviews of these topics in the literature, including quality frameworks, principles and guidelines (Tam & Clarke, 2015; Pfeffermann, 2015; Japec, et al., 2015; Lohr & Raghunathan 2017; UNECE; Eurostat 2018; Eurostat 2020; de Waal, et al., 2020).

For some NPD sources such as cash register data, the integration into official statistics is already in place and a large number of pilot studies has been conducted regarding other sources (Salvatore, 2023; Liao & Biemer, 2025). But the step from research or pilot study to production at the statistical office has often proved to be a long and slow one (Holmberg, 2025). Part of the explanation for this can be found in the conglomeration of practical, technical and legal issues that need to be addressed when new data are to be incorporated. But another hurdle is that for some of the challenges that arise in dealing with non-probability data, the statistical office lacks direct access to methods and tools suitable for the task under the strongly regulated, time-scheduled and resource-bounded conditions that the production of official statistics presents (Beaumont & Rao, 2021).

Although given much attention in research, one such challenge is undercoverage of non-probability data with respect to the population of interest. For example, various aspects of the economic statistics for Sweden could be improved with the use of data from the annual accounting reports that companies are obliged to provide. To utilize this data in a cost-effective manner requires however that the companies provide their annual reports in digital form with labeled variables that makes it doable to put the data in a database. As of 2025 about 60% of all eligible companies did this with a bias towards newer and smaller companies, putting any statistics based sole on accounting data for these more computer-savvy companies at risk of selection bias. For most applications, this has put usage of the data on hold in the hope that future legislation or other factors will solve the problem.

The risk of selection bias has been given considerable attention in the literature and a fruitful approach seems to be to combine nonprobability data with data collected from a probability sample. The flora of such estimators includes imputation methods (Yang & Kim, 2018; Kim, Park, Chen & Wu, 2021; Yang, Kim & Hwang, 2021), response propensity estimators (Lee, 2006; Rivers, 2007; Lee & Valliant, 2009; Brick, 2015; Holmberg, Ang, Clark & Loong, 2024), weight calibration (Kim, 2022; Golini & Righi, 2024) and the double robust estimator (Kim & Wang, 2019, Chen, Li, & Wu, 2020; Burakauskaitė & Čiginas, 2023). Which of these is most suitable for application depends on the conditions under which the survey is conducted. A drawback is however that these estimators all depend on assumptions such as missing at random on the non-probability data, i.e. that the missingness is subject to an unknown but adjustable selection mechanism. In practice, such an assumption is rarely satisfied, and even if it was, it may be difficult or even impossible to verify whether it holds or not. Kim and Tam (2021) proposed the Data Integration estimator which integrate the nonprobability data in the design weights, thereby reducing the assumptions made about the nonprobability data. We shall explore this estimator further in Section 3.

If the goal of using non-probability data in a survey is a reduction of costs or respondent burden, and if complementing data is to be collected from a probability sample, then this sample needs to be small. Given that auxiliary information is available a reflexive reaction might be to reach for the GREG estimator. However, as demonstrated in e.g. Lundy and Rao (2022) and Goga and Haziza (2023), for small sample sizes and/or numerous auxiliary variables, GREG might fail to reduce the variance compared to the HT estimator, produce unstable weights or even no estimates at all if the regression matrix involved becomes singular. There are sometimes workarounds for these issues, but they can be time consuming, costly and might limit the amount of auxiliary information that can be used.

The use of machine learning for variance reduction in survey sampling is a fairly new and exciting area of research (Ranalli, 2025). Lundy and Rao (2022) demonstrated that having a wider range of models to choose from can come with benefits such as greater variance reduction and a reduced dependence on properties of the auxiliary information. If the model can sort out the data itself, a higher degree of automation will be possible. The move from linear regression to machine learning does however increase the risk of overfitting of the model to the data available (Ranalli, 2025). Although the following example is intentionally exaggerated, it serves to illustrate the underlying point. Training an unrestricted regression tree model on a sample might very well yield a model that can exactly predict the target variable on the sample, without any guarantees on performance outside of the sample. A variance estimator based on the difference between the target variable and its prediction on the sample, as those usually used for GREG, will then be constantly zero, which in most cases is absurd. One solution for this could be to force restrictions on the model making over fitting harder, but this could require variable specific actions and lead to less effective estimator. A better approach, followed in this paper, is to separate the data on which the model is trained from the data where the variance is estimated (Rein, 2025).

In this paper we present and evaluate a sampling strategy that we believe can enable the use of more non-probability data in the estimation process. The estimator is based on a particular form of the data integrated estimator by Kim and Tam (2021), although it provides a further variance reduction by using an arbitrary machine learning model and externally available auxiliary information. The sampling strategy requires the availability of auxiliary information at the population level before sampling, and the presence of an identifier in the non-probability data making it possible to match it to other data. These conditions are satisfied by many both existing and up-coming data sources at several NSOs. The estimator is constructed using components that should be familiar to anyone with a background in design-based survey sampling. It is design-unbiased and we propose a design-unbiased variance estimator which is suitable for any choice of model, making it very safe to use. We demonstrate the proposed estimator with a random forest model since it usually performs well without the need of tuning hyperparameters, hence it is suitable for the industrialized conditions under which official statistics is produced. It is however possible to apply any arbitrary model to the estimator.

The layout is as follows, in Section 2 we present a short discussion to what we call some traditional estimator. This section should enable the reader to better understand the upcoming proposed sampling strategy. In Section 3 we discuss the Data Integration estimator proposed by Kim and Tam (2021). This is subsequently extended by Section 4 where we introduce and discuss our proposed sampling strategy as well as its properties, we call it the Model Assisted Data Integration (MADI). Section 5 illustrates MADI with two simulations using empirical data from Statistic Sweden. In this illustration we verify empirically that the proposed sampling strategy produce unbiased estimates and that the variance is lower compared to traditional survey estimators even for highly restricted sample sizes where the GREG estimator struggles to perform. We also demonstrate that the sampling strategy performs well with a highly restricted sample size. Lastly, we discuss some future development and suggestions moving forward in Section 6.

## 2. Traditional estimators

Let $U$ be a finite population with labels $\{1,2,\ldots,i,\ldots,N\}$ and let $(x_i, y_i)$ be a vector of auxiliary variables $x_i$ assumed here to be available for each $i \in U$ and the value of a study variable $y_i$ associated with the $i$th unit. We shall consider the estimation of the population total $t_y = \sum_U y_i$. From $U$ we draw a probability sample $s$ of size $n$. Let $\pi_i = p(i \in s)$ be the associated inclusion probability for $i \in U$ and $\pi_{ij} = P(i,j \in s)$ be the second order inclusion probability where we assume that $\pi_{ij} > 0 \ \forall i,j \in U$.

In the upcoming subsections we shall cover some basic concepts that our proposed estimator relies on. These should be rather familiar to any survey statistician, but we repeat them since they provide some pedagogical benefits for the upcoming proposed estimator. A more in-depth description of the methods described in this section can be found in e.g. chapter 3.7 and 6.3 in Särndal et al. (1992).

### 2.1 Horvitz-Thompson and Generalized Regression

We use the phrase traditional designs in this paper to refer to any probability design commonly used in survey statistics which draws a sample $s$ directly from $U$. We shall combine traditional design with the Horvitz-Thompson (HT) estimator and the Generalized Regression (GREG) estimator. We are aware of the limitations of the HT estimator but choose to include it as a reference point. The GREG estimator is commonly used in survey statistics. It is well established that the GREG estimator is biased, i.e. $E(\hat{t}_{greg}) \neq t_y$. This bias is however mitigated if the sample is large enough since it is design consistent. In practice, this leads to a minuscule bias if the sample size is reasonably large (Särndal et al., 1992). Furthermore, the variance estimator can have substantial negative bias when the number of explanatory variables is large in comparison to the size of the sample, especially in smaller samples (Lundy & Rao, 2022; Dagdoug, Goga & Haziza, 2023). Thus, variable selection may be required to ensure an accurate variance estimate, which in turn may lead to increased variance due to reduced information captured by the model. Consequently, the precision of the variance estimate may paradoxically coincide with increased variance. For more detail on the HT and GREG estimators we refer the reader to Särndal et al., (1992).

### 2.2 Difference estimator

The difference estimator is the blueprint for the GREG estimator that is less complex mathematically. This estimator utilizes auxiliary information but instead of estimating a linear model based on the probability sample, it is assumed that the linear coefficients are known beforehand for all $i$ in $U$. Let $y_i^0$ be the predicted $y_i$ value for individual $i$ using the known coefficients. The difference estimator of $t_y$ is defined as

$$\hat{t}_{y,dif} = \sum_{i \in U} y_i^0 + \sum_{i \in s} \frac{D_i}{\pi_i} \qquad (1)$$

where $D_i = y_i - y_i^0$. The main idea is to have the proxy $y_i^0$ being as close as possible to the true $y_i$ and thereby yielding an estimator with a low variance. It is important to note that (1) is unbiased for $t_y$ since $y_i^0$ does not depend on the $s$.

$$E(\hat{t}_{y,dif}) = E\left(\sum_{i \in U} y_i^0 + \sum_{i \in s} \frac{y_i - y_i^0}{\pi_i}\right) = E(y_i^0 + t_y - y_i^0) = t_y. \qquad (2)$$

This is true for any $y_i^0$, for example $y_i^0$ can be replaced with a constant and the estimator will still be unbiased, the only issue would be a potentially increased variance. If $y_i^0 = 0$ then $\hat{t}_{y,dif}$ become the HT estimator $\hat{t}_{y,HT}$.

The variance of the difference estimator is given by

$$V(\hat{t}_{y,dif}) = \sum_{i \in U} \sum_{j \in U} \Delta_{ij} \frac{D_i}{\pi_i} \frac{D_j}{\pi_j} \qquad (3)$$

where $\Delta_{ik} = \pi_{ij} - \pi_i \pi_j$. An unbiased variance estimator is given by

$$\hat{V}(\hat{t}_{y,dif}) = \sum_{i \in s} \sum_{j \in s} \frac{\Delta_{ij}}{\pi_{ij}} \frac{D_i}{\pi_i} \frac{D_j}{\pi_j}. \qquad (4)$$

The variance of the difference estimator will be low when $y_i^0$ is close to the $y_i$.

### 3. The data integration estimator

Let $A$ be a subset of $U$ of size $N_A$. In this paper we assume that the study variable $y_i$ is known for all $i \in U$ but we make no assumptions on the selection process of $A$. Thus $A$ is NPD and no design weights are available. Let $B$ be the complement of $A$ in $U$, with size $N_B$. The subsets $A$ and $B$ can be viewed as two different strata where $y_i$ is known for all units in $A$ and unknown for all units in $B$.

In this scenario, the known values of the study variable are utilized by Kim and Tam (2021) to make design-based estimates. Their idea is to estimate a parameter from observations on a probability sample $s$ drawn from $U$ using the known $y_i$-values in $A$ as auxiliary information. To do so they propose the Data Integration (DI) estimator. Let $\delta$ be the indicator function for $A$, i.e. $\delta_i = 1$ for $i \in A$ and 0 otherwise. Then the DI estimator is defined as a form of post-stratification estimator

$$\hat{t}_{y,DI}^{K\&T} = t_A + N_B \frac{\sum_{i \in s} \pi_i^{-1} y_i (1-\delta_i)}{\sum_{i \in s} \pi_i^{-1}(1-\delta_i)} \qquad (5)$$

where $t_A = \sum_{i \in A} y_i$ and the subscript $K\&T$ refers to the version of the DI estimator proposed by Kim and Tam. The estimator (5) is design consistent but not design unbiased in general. Note that the randomness of the probability sample only comes into play in the second term of the estimator and this term doesn't involve units from $s \cap A$. Kim and Tam (2021) further develop (5) to deal with the cases where unique identifiers for the units in $A$ are predictable but not explicitly known and when the $y$ variable is measured with measurement errors. This is the reason why they want $s$ to contain units from both $A$ and $B$ and it makes it possible to incorporate NPD in an ongoing survey without having to adjust the design. Since their estimator is not linear, variance expression is approximated with a Taylor linearization.

At Statistics Sweden and most other counties in northern Europe nonprobability datasets with known identifiers are not uncommon. With known identifiers and when measurement errors in $A$ isn't an issue, it seems rational to draw the sample solely from $B$. Let $s_b$ be a probability sample drawn from $B$ with inclusion probabilities $\pi_{iB}$ and second order inclusion probabilities $\pi_{ijB} > 0$. In other words, this is a traditional probability sample drawn from the subsection of $U$ not covered by the $NPD$. The estimator (5) then simplifies to

$$\hat{t}_{y,DI}^{HT} = t_A + \sum_{i \in s_b} \pi_i^{-1} y_i \tag{6}$$

where $\sum_{i \in s} \pi_i^{-1} y_i$ is a HT estimator based on the part of the probability sample. The estimator (6) is design unbiased and encompasses a better sample economy since the whole sample is used in the second term. The variance of (6) is

$$V(\hat{t}_{y,DI}^{HT}) = V(t_{y,A}) + V(\hat{t}_{y,B}) = 0 + V(\hat{t}_{y,B}) = \sum_{i \in B} \sum_{j \in B} \Delta_{ij} \frac{y_i}{\pi_i} \frac{y_j}{\pi_j}. \tag{7}$$

An unbiased variance estimate is

$$\hat{V}(\hat{t}_{y,HT}^{B}) = \hat{V}(t_{y,A}) + \hat{V}(\hat{t}_{y,B}) = 0 + \hat{V}(\hat{t}_{y,B}) = \sum_{i \in s_b} \sum_{j \in s_b} \frac{\Delta_{ij}}{\pi_{ij}} \frac{y_i}{\pi_i} \frac{y_j}{\pi_j}. \tag{8}$$

The second part of (6) can be replaced with any traditional survey-based estimator and the variance will inherit the properties of the selected estimator. For example, if $\sum_{i \in s_b} \pi_i^{-1} y_i$ in (6) is replaced with a GREG estimator, the corresponding variance will be approximated with the Taylor linearization.

### 4. A Model Assisted Data Integration survey strategy

The MADI survey strategy proposed and evaluated in this paper takes it starting point in the data integration estimator by Kim and Tam (2021) described in Section 3. We assume that the subsets $A$ and $B$ form a partition of $U$ and that the study variable $y_i$ is known before sampling for units in $A$. We assume that a vector $x_i$ of auxiliary information is available for each $i \in U$ and that there is a known and unique identifier for each unit in $U$ making it possible to connect the known $y$ values and the auxiliary vectors to $U$ without matching errors. We can write (6) in a more general way

$$\hat{t}_{y,DI} = t_A + \hat{t}_B \tag{9}$$

where $\hat{t}_B$ can be any estimator of $t_b$.

Let $\mu(x, A)$ be a model trained on $A$ based on the observed study variable and the known auxiliary information $x_i$ from units in $A$. No assumptions are made about the structure of $\mu(x, A)$, it can be any arbitrary model. The model $\mu(x, A)$ can be used to make predictions $\mu(x_i, A)$ of $y_i$ from $x_i$ for each $i \in U$. Replacing $\hat{t}_B$ in (9) with a difference estimator (1) based on the predicted $y$-values results in the estimator

$$\hat{t}_{y,MADI} = \sum_{i \in A} y_i + \sum_{i \in B} \mu(x_i, A) + \sum_{i \in s} \frac{e_i}{\pi_i}, \tag{10}$$

where $e_i = y_i - \mu(x_i, A)$. Since $s_b$ is a probability sample, we can aggregate it using the design weights $\pi_{iB}$ to estimate the population total. In machine learning literature it is common to use a partition the available data into a training and testing test, where the model $\mu(x)$ is developed solely with the training set and evaluated on the test set. We employ a similar mindset in (10) where $\mu(x, A)$ is trained on the NPD set and evaluated on the probability sample $s_b$. Potential errors introduced by the model is corrected for in the third term of (10). This is fundamentally different to e.g. the GREG estimator where the model is evaluated on residuals produced by the sample, which is the reason for the biased introduced by the estimator. In words this estimator combines the total of the known $y_i$ in $A$, a model estimate for the total of $y_i$ in $B$, and a third term that estimates the error in the model term. Since $\mu_{MADI}(x_i, A)$ does not depend on a probability sample we can use the theory from (2) to show that (10) is unbiased

$$E(\hat{t}_{y,MADI}) = E\left(\sum_{i \in A} y_i + \sum_{i \in B} \mu(\boldsymbol{x_i}, A) + \sum_{i \in s_b} \frac{e_i}{\pi_{iB}}\right) = t_{y,A} + t_{\mu,B} + t_{y,B} - t_{\mu,B} = t_y$$

where $t_{\mu,B}$ correspond to the total of $y$ in $B$ based on the predictions from $\mu(x, A)$. By similar reasoning as shown in (7), the variance for the proposed estimator is given by

$$V(\hat{t}_{y,MADI}) = \sum_{i \in B} \sum_{j \in B} \Delta_{ijB} \frac{e_i}{\pi_{iB}} \frac{e_j}{\pi_{jB}}.$$

where $\Delta_{ijb} = \pi_{ijB} - \pi_{iB}\pi_{jB}$, and an unbiased estimator for the variance is given in a similar way as in (8) by

$$\hat{V}(\hat{t}_{y,MADI}) = \sum_{i \in b} \sum_{j \in b} \frac{\Delta_{ijB}}{\pi_{ijB}} \frac{e_i}{\pi_{iB}} \frac{e_j}{\pi_{jB}}. \tag{11}$$

The proposed estimator (10) mitigates the risks associated with overfitting by employing a machine learning mindset to the sampling strategy. The risk associated with a poorly fitted model is an increased variance, but since both the estimator and the proposed variance estimator are unbiased, in practice this issue will be revealed by the variance estimate. As we shall see in Section 5.2, empirical results indicate that (10) is rather robust even against poorly fitted models.

It is also important to note that since $\mu(\boldsymbol{x_i}, A)$ is trained on the NPD, which perceptually is much larger compared to a traditional probability sample, the risk of having too few data points for a given model is mitigated. For example, a neural network model with several dimensions is most often not possible to use in traditional surveys since the number of required observations is usually too few. Furthermore, it also allows for more extensive use of auxiliary information in the modeling process. As illustrated by e.g. Lundy & Rao, 2022, there is a limit to the number of possible auxiliary variables the model can handle in relationship to the number of sampled units. Under the assuming that the NPD is of decent size, this issue is counteracted. This will be further illustrated in the simulations in this paper.

The proposed $MADI$ leverages the beneficial properties of the difference estimator to ensure an unbiased estimator. This approach allows for flexibility when selecting a model since the theory of the difference estimator will ensure that the estimator is unbiased. For someone with a background in survey statistics, this estimator should be rather easy to understand even if the conceptually mindset differs slightly from the traditional survey approach.

The $MADI$ sampling strategy requires both a design and an estimator which differs from Kim and Tam (2021) which only propose an estimator. Thus, (5) can be incorporated in an already ongoing production without having to adjust the design. Consequentially, the threshold to explore $MADI$ may be a bit higher but as we shall see in the simulations, the potential improvements in variance reduction may persuade practitioners that the approach is justified. Possibly most important part of the $MADI$ is the vastly reduces sample size needed in the probability sample compared to traditional estimators. Since the proposed sampling strategy have a targeted focus on the subset of $U$ which we are interested in, the potential reduction can be quite large.

## 5. Simulations

This section illustrates the proposed $MADI$ sampling strategy described in Section 4 by a simulation with data from Statistics Sweden. The example aims to demonstrate (i) the validity of $MADI$ and (ii) the efficiency of the $MADI$ compared with other design-based methods.

The simulation is based on data from the Swedish Income and Taxation (IaT) register from year 2021. The register includes variables related to different types of income, i.e., income related taxes, pensions, and social insurance benefits. For computational reasons, we constructed a population by drawing a random subset of size $N = 9\,838$, excluding extreme outliers. The study variable $y$ is Annual Determined Earned Income measured in Swedish Krones is defined as

*"Income from employment plus income from business activities minus general deductions."* (SEB, 2025).

We selected twelve explanatory variables, where $X_1$ represents age, $X_2$ represents gender, and $X_3$ to $X_{12}$ are associated with taxes. In Figure 1, we illustrate the relationship between each explanatory variable and the study variable.

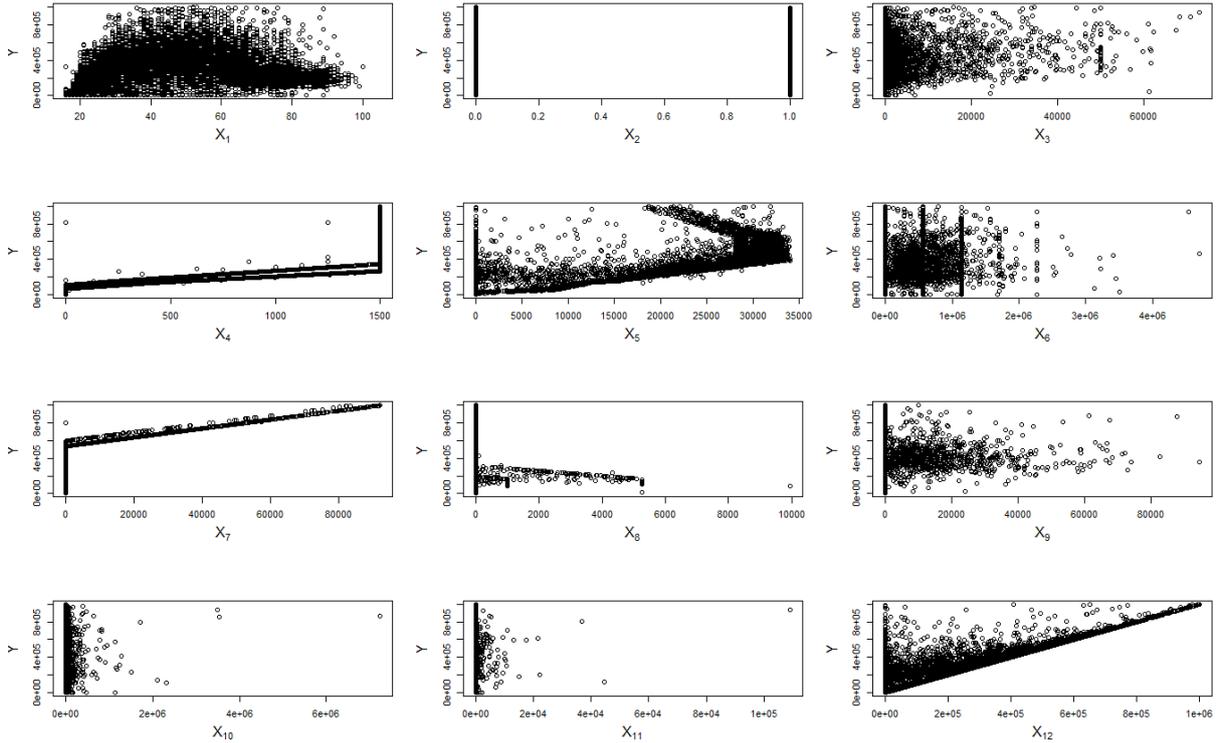

Figure 1. The relationship between the study variable and the explanatory variables for the population U illustrated by a scatter plot for each variable.

The nonprobability dataset was generated directly from $U$ and remained fixed for all estimators. In Simulation 1 the NPD were of size $0.7N$ and in Simulation 2 we elaborate of difference proportions of sizes of the NPD. The process of how the NPD was generated is explained in more detail in each of the two simulation sections below.

Two different designs are used in the simulation, denoted $SRS_U$ and $SRS_B$. In both cases we use a simple random sample, the difference between the two is that in $SRS_U$ the sample is drawn from $U$ while in $SRS_B$ the sample is drawn from $B = U \setminus A$. In $SRS_U$ we draw a sample of size $n$ where each unit has the inclusion probability $\pi_i = \frac{n}{N}$. This design represents a traditional survey where an estimate is calculated based on a sample drawn from a population. Similarly, in $SRS_B$ a SRS of size $n_b$ is drawn with inclusion probabilities $\pi_{iB} = \frac{n_b}{N_B}$. This design is used to illustrate the scenario where we have information about the study variable in an incomplete register or where we have access to other nonprobability data that do not cover the whole population.

We employ the following sampling strategies in the simulations

$SRS_U - HT$: Simple random sample from $U$ coupled with the HT-estimator
$SRS_U - GREG$: Simple random sample from $U$ coupled with the GREG estimator
$SRS_U - RF$: Simple random sample from $U$ coupled with a naïve RF estimator
$SRS_B - DI_{HT}$: Simple random sample from $B$ coupled with the $DI$-estimator and a HT-estimator
$SRS_B - DI_{GREG}$: Simple random sample from $B$ coupled with the $DI$-estimator and GREG estimator
$SRS_B - DI_{RF}$: Simple random sample from $B$ coupled with the $DI$-estimator and a naïve RF estimator
$SRS_B - MADI_{OLS}$: Simple random sample from $B$ coupled with a $MADI$-estimator and an OLS model
$SRS_B - MADI_{RF}$: Simple random sample from $B$ coupled with a $MADI$-estimator and a RF model

The naïve RF estimator is a model based on the GREG estimator described in Section 2.2, but where we replace the WLS model with a random forest model. This is inappropriate, given that nonparametric models lack identifiable parameters, a limitation that can easily translate into biased estimates, since the estimate will rely on sample specific prediction patterns rather than a stable representation of the underlying relationship, hence we call it naïve. There are ways to alter the GREG estimator which allows the use a RF or other nonlinear models while maintaining design-unbiasedness, e.g. the LLO-estimator which utilize a subsampling Rao-Blackwellisation method proposed by Sande and Zhang (2021). Initial testing did however show that such an estimator come at a high computational cost and a slightly higher variance compared to a traditional GREG. Since we want to focus on minimizing the variance in combination with using an estimator which is easy to use, we do not include the LOO-estimator.

**5.1 Simulation 1: Comparing the performance of different estimators**

Our aim is to illustrate that $MADI$ performs well even when a missing not at random mechanism has generated the NPD. A response propensity variable $\theta_i$ was generated which can be seen as the probability for unit $i$ being part of $A$. The variable $\theta_i$ was generated from the uniform distribution where the lower limit was 0 and the upper limit was based on the quantiles of the study variable ranging from 0.1 for individuals with the lowest values on the study variable up to 1 for the individuals with the highest values on the study variable. This resulted in $cor(y, \theta) = 0.599$. This positive correlation can be interpreted as a selection bias in $A$ where higher values on the study variable generally increased the probability of being in the nonprobability dataset; see Figure 2.

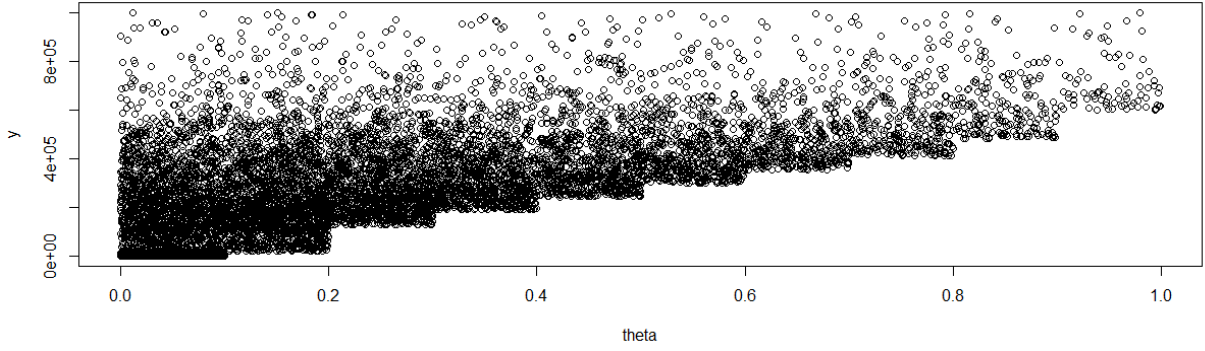

*Figure 2. The study variable on the Y axis and the simulated probability of being included in $A$ on the X axis.*

The process of generating units to $A$ was as follows, we randomly selected a unit $i$ from $U$ and generated a $Uni\sim(0,1)$ variable. If the generated uniform variable value was lower than $\theta_i$, the unit was selected to $A$. This process was then repeated for a new random unit in $U \notin A$ until 70% of the population was allocated to $A$ and the remaining 30% of the population was allocated to $B$. Among the units in $A$, the mean of the study variable was 350 037 SEK while the mean among the units in $B$ were 206 584 SEK. The population and NPD remained fixed in all simulations.

The same sample size was used regardless of sample strategy. We started with a sample size of $n = n_b = 25$ and gradually increased it with 25 more units for each new iteration until we reached $n = n_b = 800$. Moving forward, when the size of $n$ is described, it is implied that the same size used for $n_b$.

The simulation has the following structure:

1. For each sample size $n_k, k = 25, 50, \ldots, 800$ and each design $SRS_U, SRS_B$, $M$ samples $s_m$ are drawn where $m = 1, \ldots, M$. We set $M = 1000$.
2. For each sampling strategy, the population total $\hat{t}_y$ and the variance $\hat{V}(\hat{t}_y)$ are estimated.
3. The bias was approximated with $B(\hat{t}_y) = \frac{1}{M}\sum_{m=1}^{M}(\hat{t}_{y,m} - t_y)$, where $t_y$ is the true population total.
4. Similarly to the bias, the mean squared error was approximated by $MSE(\hat{t}_y) = \frac{1}{M}\sum_{m=1}^{M}(\hat{t}_{y,m} - t_y)^2$.
5. The coverage probability is calculated as the proportion of confidence intervals which covers the true total, where the confidence interval for each estimate is $CI(\hat{Y}_m) = \hat{t}_{y,m} \pm t_{n-1,1-\frac{\alpha}{2}}\hat{V}(\hat{t}_{y,m})^{0.5}$. We use the $t$-distribution with $\alpha = .95$ since some of the sample sizes are quite small. This approach is also in line with the recommendations for the GREG estimator with multiple auxiliary variables from Särndal et al. (1992) chapter 7.9.1.

The RMSE is presented in Table 1 for each sampling strategy. In some cases, it was not possible to calculate the point estimate for all $n_k$ for a specific estimator, which is indicated by NA. The GREG estimator is ill suited for small sample sizes combined with a vast amount of auxiliary variables since it can make the matrix calculations singular. The proposed estimator (10) combined with the OLS model does not suffer from this issue since the model is based on the large data provided in $A$.

We observe that the $SRS_B - MADI_{RF}$ yeild the lowest RMSE regardless of sample size compared to all other sampling strategies. The $SRS_B - MADI_{OLS}$ has the second lowest RMSE for small sample sizes. In this particular setup, $SRS_B - DI_{GREG}$ require a sample size of $n = 400$ in order to consistently produce an estimate. At this sample size, the difference in between $SRS_B - MADI_{OLS}$ and $SRS_B - DI_{GREG}$ is quite substantial. This difference does however degrade as $n$ increases. This is reasonable since as the amount of information in the probability sample increases, the underlying $WLS$ model in $SRS_B - DI_{GREG}$ receive more information while the information in the $OLS$ remain constant for $SRS_B - MADI_{OLS}$.

The $SRS_B - DI_{RF}$ perform relatively well in terms of RMSE, but as we shall soon see, the bias of the estimator provides a grimmer picture. The estimators that do not utilize the data from $A$ are generally outperformed by both $MADI$ and the $DI$-estimator regardless of how the model is produced.

Table 1. RMSE for each sampling strategy with different sample sizes. Each row corresponds to the sample size shown in the left most column.

| n | $SRS_U - HT$ | $SRS_U - GREG$ | $SRS_U - RF$ | $SRS_B - DI_{HT}$ | $SRS_B - DI_{GREG}$ | $SRS_B - DI_{RF}$ | $SRS_B - MADI_{OLS}$ | $SRS_B - MADI_{RF}$ |
|---|---|---|---|---|---|---|---|---|
| 25 | 398 247 810 | NA | 184 537 498 | 111 661 040 | NA | 52 417 309 | 26 160 629 | 14 140 071 |
| 50 | 263 460 971 | NA | 104 916 863 | 78 446 342 | NA | 30 161 110 | 19 207 066 | 9 689 774 |
| 75 | 230 135 875 | NA | 79 047 498 | 64 752 694 | NA | 23 754 783 | 14 979 118 | 7 781 539 |
| 100 | 203 540 913 | NA | 64 961 733 | 56 220 649 | NA | 18 583 053 | 13 013 332 | 6 544 294 |
| 125 | 180 482 612 | NA | 56 056 863 | 50 735 136 | NA | 15 657 864 | 11 827 074 | 5 959 649 |
| 150 | 158 102 469 | NA | 48 494 240 | 44 505 950 | NA | 13 418 470 | 10 642 446 | 5 739 653 |
| 175 | 142 677 306 | NA | 42 740 957 | 42 700 983 | NA | 12 532 907 | 9 477 876 | 5 028 514 |
| 200 | 144 547 280 | NA | 39 598 863 | 38 161 849 | NA | 11 230 559 | 8 671 260 | 4 617 987 |
| 225 | 137 902 795 | NA | 35 534 114 | 36 072 446 | NA | 9 701 070 | 8 407 122 | 4 499 520 |
| 250 | 124 613 488 | NA | 33 046 985 | 34 047 626 | NA | 9 135 436 | 7 289 301 | 4 013 734 |
| 275 | 115 055 068 | NA | 31 529 337 | 32 877 435 | NA | 8 343 682 | 7 606 113 | 3 838 691 |
| 300 | 110 082 980 | 98 175 476 | 28 547 842 | 30 781 534 | NA | 7 903 731 | 7 144 596 | 3 834 423 |
| 325 | 105 738 177 | NA | 26 639 275 | 29 814 624 | NA | 7 618 878 | 6 656 411 | 3 658 564 |
| 350 | 103 271 399 | NA | 26 217 208 | 27 985 470 | NA | 7 080 672 | 6 566 189 | 3 495 026 |
| 375 | 97 812 968 | 30 722 919 | 25 204 929 | 27 283 917 | NA | 6 447 833 | 6 623 387 | 3 301 842 |
| 400 | 98 282 721 | NA | 23 289 510 | 25 333 288 | 6 315 997 | 6 212 076 | 5 823 297 | 3 139 128 |
| 425 | 97 710 940 | 26 644 918 | 22 607 780 | 24 164 980 | 6 315 132 | 5 827 537 | 5 982 677 | 3 068 736 |
| 450 | 90 481 972 | 24 868 151 | 21 996 644 | 24 206 587 | 6 042 878 | 5 854 115 | 5 698 867 | 2 942 875 |
| 475 | 90 398 384 | 23 452 468 | 20 099 198 | 23 752 619 | 5 921 500 | 5 504 217 | 5 485 983 | 2 855 279 |
| 500 | 83 059 157 | 23 794 044 | 19 119 036 | 22 236 692 | 5 612 293 | 5 193 068 | 5 414 175 | 2 965 647 |
| 525 | 84 373 071 | 22 992 157 | 18 885 959 | 22 425 174 | 5 516 354 | 5 158 367 | 5 265 215 | 2 727 982 |
| 550 | 82 611 772 | 22 034 168 | 17 975 726 | 21 150 155 | 5 149 905 | 5 010 802 | 4 863 203 | 2 643 324 |
| 575 | 81 876 268 | 21 565 582 | 17 744 168 | 20 541 249 | 5 143 795 | 4 691 417 | 5 047 942 | 2 664 485 |
| 600 | 81 526 518 | 21 442 483 | 17 847 497 | 19 530 201 | 4 965 038 | 4 560 902 | 4 754 364 | 2 388 022 |
| 625 | 81 804 506 | 20 766 792 | 16 811 431 | 20 349 868 | 4 729 044 | 4 215 869 | 4 838 164 | 2 422 428 |
| 650 | 72 034 877 | 19 839 757 | 15 880 722 | 20 441 351 | 4 803 224 | 4 271 883 | 4 641 866 | 2 320 049 |
| 675 | 75 092 058 | 19 683 265 | 16 042 724 | 19 447 512 | 4 710 821 | 4 005 352 | 4 426 950 | 2 282 904 |
| 700 | 72 451 663 | 18 490 695 | 15 651 600 | 19 130 370 | 4 418 227 | 3 971 655 | 4 342 417 | 2 216 610 |
| 725 | 75 329 036 | 18 548 422 | 15 621 366 | 18 595 330 | 4 246 013 | 3 847 010 | 4 373 562 | 2 317 175 |
| 750 | 72 206 074 | 18 539 079 | 15 163 303 | 17 935 026 | 4 130 291 | 3 639 398 | 4 043 888 | 2 195 499 |
| 775 | 67 767 295 | 17 458 135 | 14 990 010 | 16 832 690 | 4 205 822 | 3 614 226 | 4 170 831 | 2 188 314 |
| 800 | 68 856 953 | 18 173 416 | 14 047 547 | 16 499 109 | 4 069 037 | 3 501 613 | 3 992 083 | 2 063 685 |

In practice, given that the conditions for the $MADI$ sampling strategy is fulfilled, a practitioner might consider one of the $DI$-estimator or MADI as a realistic choice of estimator. The table above illustrates that in this simulation MADI gives a lower RMSE than the DI-estimator, regardless of model. The RMSE of the $SRS_B - MADI_{RF}$ is about half of $SRS_B - MADI_{OLS}$ regardless of sample size in the probability sample.

In Figure 3 we illustrate the relationship between bias and sample size of the different sampling strategies. As we expect, the only estimator that shows any bias is the naïve random forest model. The $MADI$ sampling strategy can utilize the information in the nonprobability dataset well without bias even when the probability sample is small, both when using the OLS and the RF model. The bias for the $MADI$ is also quite stable, hovering around zero for all sample sizes. This can be compared to the HT estimator which is also unbiased but varies around zero in a much greater extent. The problem with the naïve RF models that introduce bias in the estimates. This is particularly evident for small sample sizes. It does however remain for all sample sizes in this simulation as shown in Figure 3.

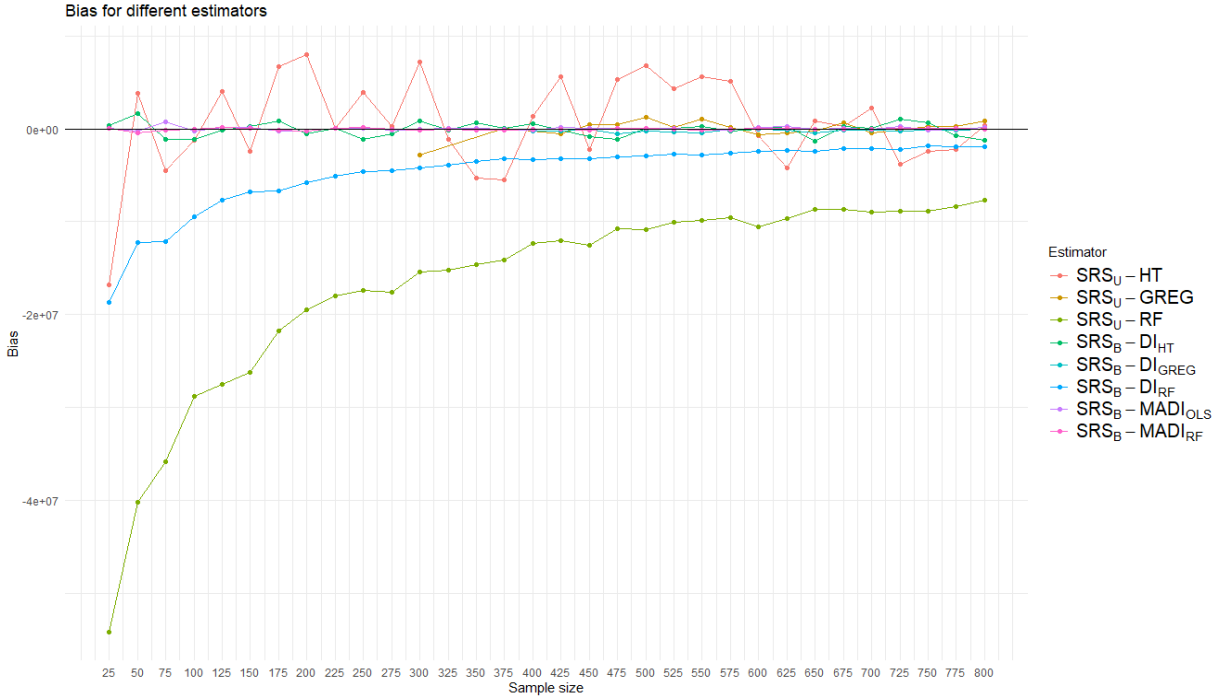

*Figure 3. Bias for each coupling of design and estimator with different sample sizes.*

In Figure 4, we illustrate the coverage probability of each coupling of design and estimator, that is the proportion of confidence intervals which cover the true value. The target is a coverage of 95%. We observe that the $MADI$-estimator, both with RF and OLS, have very good coverage. As expected, the HT estimator performs well in terms of coverage. We also observe that the GREG estimator has a coverage slightly below 95% for the samples sizes for which it can produce estimates. It does however seem to increase as the sample size increases. Since the number of auxiliary variables is quite large in relation to the sample size, this goes well in line with the conclusions in Lundy & Rao (2022). We have excluded the coverage probability for the naïve RF models since we lack a decent variance estimator.

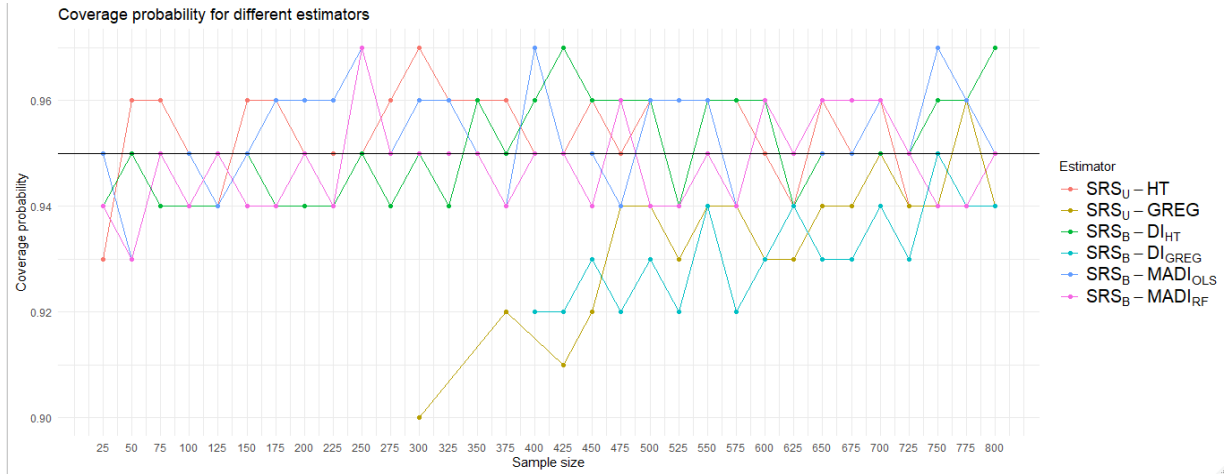

*Figure 4. Coverage probability for some couplings of designs and estimators. The Naïve random forest models are not included in this figure due to not having an appropriate variance estimator.*

### 5.2 Simulation 2: Sample size for a given coefficient of variation

In the first simulation we illustrated that the $MADI$ sampling strategy performs well in comparison to other methods when focusing on having a low RMSE. However, in many situations it might be the size of the sample for a given RMSE that is of more interest since that often yields a lower cost for the NSO. In this section we shall investigate what sample size is needed given a coefficient of variation (square roof of the variance over the parameter) of 1%. Since we have shown that the $MADI$ sampling strategy is unbiased, we chose to use the CV here instead of the RMSE since it provides a more standardized measurement.

Based on the results in the Simulation 1, we can conclude that the $SRS_B - MADI_{RF}$ is superior compared to all other methods on the dataset that we investigate. Therefore, this section aims to investigate how that specific sampling strategy performs under different circumstances. We shall compare it only to $SRS_U - HT$ and $SRS_U - GREG$, i.e. the HT and the GREG when $s$ is drawn from $U$. We have chosen those two since they best represent the estimates NSO usually use in practice.

We shall look at the same population while altering two parameters. The first parameter we will alter is the percentage of $U$ that is generated in $A$. Let subsets $A_{kl} \subseteq U$ for $k = 1,2,...,8$ and $l = 1,...,9$ be proportionally to $U$ such that $|A_{kl}| = 0.1l * |U|$, i.e. the size of $A_{kl}$ range from 10% of the population size up to 90% of the population size. The $k$ parameter shall alter the way that $A$ is generated from $U$. In Simulation 1 we aimed to have a relatively high selection bias in $A$. In this simulation we shall study how the $SRS_B - MADI_{RF}$ performs under different levels of selection bias in the NPD.

$A_{1l}$ and $A_{2l}$ represent the most extreme selection bias in this illustration. They are generated by simple allocating all units with the highest value ($A_{1l}$) or the lowest value ($A_{2l}$) to the NPD $A_{kl}$. In Figure 5 we show a histogram of $y_i$ with red lines which mark the cutoff points, the bar lines represent the deciles of the distribution of y. For $A_{1l}$, going from right to left, all units to right of the $lth$ line are included and for $A_{2l}$, going from left to right, all units to the left of the $l:th$ line are included in the NPD, the rest of the units are allocated to $B_{kl}$. In other words, it is fully deterministic whether unit $i$ is allocated to $A$ or $B$ based solely on the value of the study variable.

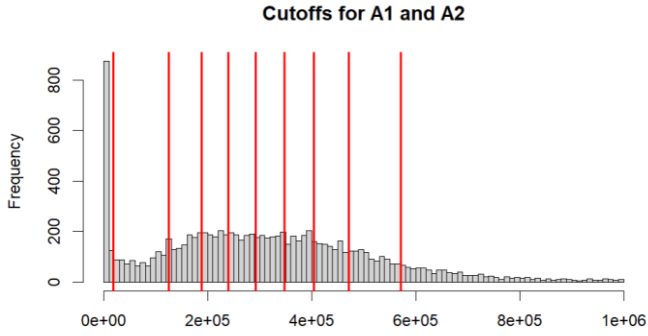

*Figure 5. Histogram of the study variable with red lines illustrating the deciles of the distribution.*

For $A_{3l}, ..., A_{8l}$ there was an element of randomness in the selection process of the NPA. For each scenario, a $\theta_{ik}$ was generated to ensure different levels of selection bias in $A_{kl}$. The process of generating $\theta_{ik}$ will be described in the next paragraphs. Once the $\theta_{ik}$ had been generated, the same process of selecting units to $A_{kl}$ were used as described in the Simulation 1.

$A_3$ is generated the same way as A was generated in Simulation 1 in Section 5.1, i.e. $\theta_{i3}$ is generated for each unit in $U$ ranging from 0 up to a limit based on their study variable. The response propensity $\theta_{i4}$ was generated as a function of $\theta_{i3}$ simply as $\theta_{i4} = 1 - \theta_{i3}$. These two methods of generating response propensities are used to illustrate a relatively high selection bias where units in $A_{3l}$ overall have higher values of the study variable compared to $B_{3l}$. The selection bias in $A_{4l}$ is designed to instead have units with lower values on the study variable compared to the units in $B_{4l}$. In both these scenarios, $\theta_i$ is generated in such a way that it has high spread of high values of the study variable and low spread for units with low values on the study variable. In Figure 6 below we illustrate a scatter plot between the response propensities $\theta_{i3}$ and $\theta_{i4}$ respectively and the study variable.

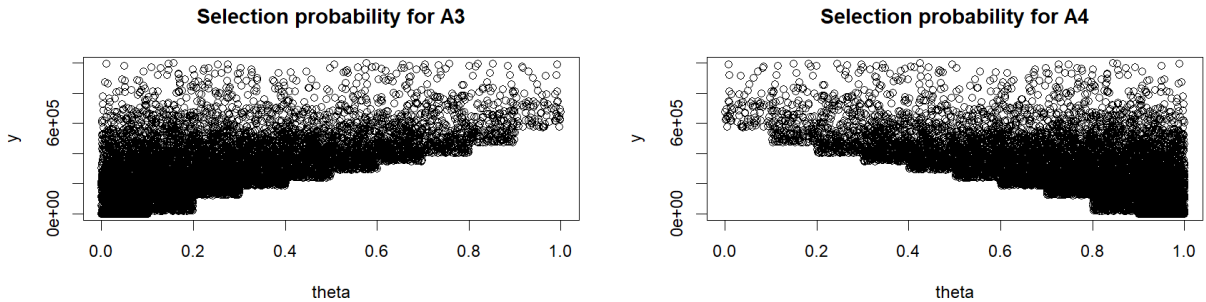

*Figure 6. The left scatter plot shows the relationship between the study variable and $\theta_{i3}$. The right figure illustrates the relationship between the study variable and $\theta_{i4}$.*

For $A_{5l}$ and $A_{6l}$ we generate a response propensity $\theta_{ik}$ to have a V-shaped pattern to $y_i$, see Figure 7. This was done by generating $\theta_{i5}$ as a function of a standardized $y_i$ and limit it to ranging between 0 and 0.95 with some additional noise added. Additionally, we randomly drew 5% of the population and set their $\theta_{i5} = 0$ to illustrate a Missing Not At Random (MNAR) situation. In addition to the MNAR, this scenario illustrates a situation where units with high and low values on the study variable are more likely to be included in $A_{5l}$ while units with average values on the study variable are much less likely to be in the NPD. We also generated $\theta_{i6}$ as a direct function of $\theta_{i5}$ by $\theta_{i6} = 1 - \theta_{i5}$, with the exception for the units with $\theta_{i5} = 0$ who remained at $\theta_{i6} = 0$. Thus, $A_{6l}$ illustrate MNAR as well as the opposite compared to $A_{5l}$ where units with average values on the study variable are much more likely to be included in the NPD while units that are either on the higher or lower spectrum of $y_i$ are less likely to be included.

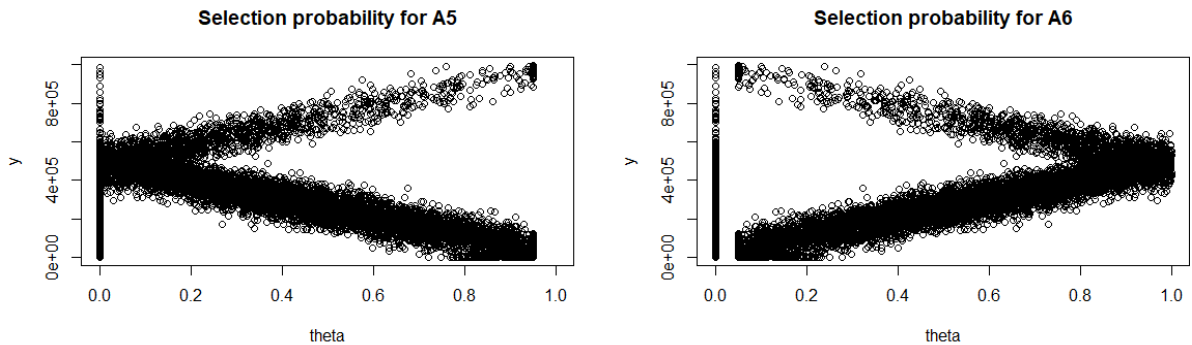

*Figure 7. The left scatter plot shows the relationship between the study variable and $\theta_{i5}$. The right figure illustrates the relationship between the study variable and $\theta_{i6}$.*

For scenario $A_{7l}$ we generated $\theta_{i7}$ as a random uniform variable with lower limit 0 and an upper limit as a function of the standardized $y_i$ where low values allowed for a high spread and high values allowed for a lower spread, see Figure 8. In this scenario, high values of the study variable were very unlikely to be in $A_{7l}$ and low values were very likely to be included in $A_{7l}$. Similarly to previous illustrations we generated $\theta_{i8} = 1 - \theta_{i7}$ for $A_{8l}$, thus giving a scenario where high values were likely to be included in $A_{8l}$ and low values were unlikely to be included. The two scenarios here are quite similar to $A_{3l}$ and $A_{4l}$ with the difference being where $\theta_{ik}$ have a high or low spread.

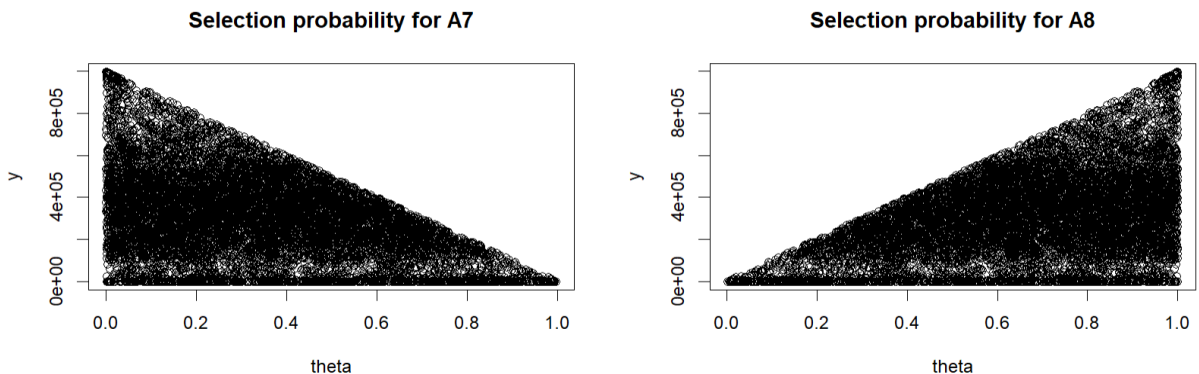

*Figure 8. The left scatter plot shows the relationship between the study variable and $\theta_{i7}$. The right figure illustrates the relationship between the study variable and $\theta_{i8}$.*

In Table 2 below we illustrate the mean value of the study variable $y_i$ in $A_{kl}$ and $B_{kl}$ for all possible $k$ shown by rows and $l$ illustrated on the columns. This should give an indication of the level of selection bias in each scenario. We observe an exceptionally high difference in mean values between $A_{kl}$ and $B_{kl}$ for $k = 1,2$ which is expected since these scenarios use cutoff points based on the study variable. The difference in the other scenarios is less pronounced but still visible for all $A_{kl}$.

*Table 2. Each cell shows the mean of $y_i$ for $A_{kl}$ and $B_{kl}$ where $l$ is represented in different columns as the proportional size of $A_{kl}$ from $U$ and $k$ is represented by rows indicating the selection process of $A_{kl}$.*

|          | 10%    | 20%    | 30%    | 40%    | 50%    | 60%    | 70%    | 80%    | 90%    |
|----------|--------|--------|--------|--------|--------|--------|--------|--------|--------|
| $A_{1l}$ | 702818 | 609254 | 551557 | 507715 | 470249 | 436196 | 404410 | 373785 | 340951 |
| $B_{1l}$ | 263018 | 231426 | 202169 | 173212 | 143764 | 113206 | 79799  | 39964  | 2606   |
| $A_{2l}$ | 2524   | 39964  | 79799  | 113206 | 143764 | 173212 | 202169 | 231426 | 263018 |
| $B_{2l}$ | 340769 | 373785 | 404410 | 436196 | 470249 | 507715 | 551557 | 609254 | 702818 |
| $A_{3l}$ | 402603 | 399047 | 397125 | 381699 | 373889 | 364999 | 352552 | 338558 | 323371 |
| $B_{3l}$ | 296383 | 283991 | 268392 | 257216 | 240124 | 220012 | 200714 | 180836 | 159766 |
| $A_{4l}$ | 263877 | 273093 | 267340 | 272268 | 273918 | 275202 | 276593 | 282424 | 288113 |
| $B_{4l}$ | 311800 | 315487 | 324003 | 330164 | 340095 | 354718 | 377987 | 405313 | 477015 |
| $A_{5l}$ | 226549 | 232548 | 230905 | 238657 | 242258 | 252213 | 263167 | 278913 | 297507 |
| $B_{5l}$ | 315949 | 325626 | 339616 | 352569 | 371756 | 389205 | 409319 | 419355 | 392485 |
| $A_{6l}$ | 386215 | 383985 | 376508 | 378903 | 371579 | 364791 | 357929 | 343788 | 320649 |
| $B_{6l}$ | 298204 | 287757 | 277226 | 259080 | 242435 | 220324 | 188166 | 159918 | 184252 |
| $A_{7l}$ | 254864 | 257267 | 259471 | 260613 | 264043 | 273180 | 276490 | 286037 | 294922 |
| $B_{7l}$ | 312802 | 319445 | 327375 | 337934 | 349971 | 357752 | 378226 | 390863 | 415744 |
| $A_{8l}$ | 322552 | 339419 | 340915 | 331382 | 335557 | 330436 | 329925 | 326606 | 320559 |
| $B_{8l}$ | 305279 | 298902 | 292478 | 290758 | 278457 | 271860 | 253521 | 228630 | 185066 |

The coefficient of variation (CV) is calculated as a function of the theoretical variance. The sample size needed for each strategy to achieve a CV of 1% for each $A_{kl}$ is presented in Table 3. The formula to calculate the sample size needed for each estimator is

$$n_{ht} = \frac{N^2 S_y^2}{(YCV_{target})^2 + NS_y^2} \tag{12}$$

$$n_{GREG} = \frac{N^2 S_{e,GREG}^2}{(YCV_{target})^2 + NS_{e,GREG}^2} \tag{13}$$

$$n_{MADI} = \frac{N_B^2 S_{DI}^2}{(Y_B CV_{target})^2 + N_B S_{DI}^2} \tag{14}$$

where $CV_{target} = 0.01$. Note that the sample size needed for $SRS_B - MADI_{RF}$ is based on the population size in $B$.

The first two rows in Table 3 illustrate $SRS_U - HT$ and $SRS_U - GREG$. Since these estimates do not depend on $A_{kl}$, they are constant over $l$ and $k$. The $SRS_U - HT$ requires 3044 sampled units and the $SRS_U - GREG$ requires 269 sampled units. The table further shows the $SRS_B - MADI_{RF}$ for each $A_{kl}$ where $k$ are presented on different rows and $l$ are presented on columns. We observe that for $k = 3,…,8$, $SRS_B - MADI_{RF}$ outperform both the $SRS_U - GREG$ and $SRS_U - HT$ for all sizes of $l$ except for $A_{68}$ in this illustration. It may seem counter intuitive that the sample size needed is not strictly decreasing as $l$ increases. In a survey context, we are used to have a fixed $N$ in the formula. However, the size of $N_B$ decrease as $l$ increase since it depends on the size of $A_{kl}$. Since $N_B$ is squared in the numerator of (14) but not in the denominator, the effect of this change can have a negative impact on the required sample size. Two scenarios can occur here which lead to an increase in required sample size. If $N_B$ decrease but $Y_B$ remain constant or only have a miniscule reduction, the number of sampled units needed increases assuming $S_{DI}^2$ is constant. The second scenario is that $S_{DI}^2$ is constant or even increases as $N_B$ increases which also lead to an increased number of sampled units needed assuming $Y_B$ remains constant. This is most pronounced in $A_{6l}$ where units with an average $y_i$ have a higher inclusion probability compared to the tail values of the study variable.

The other situation where the $SRS_U - GREG$ requires a lower sample size than $SRS_B - MADI_{RF}$ is when the selection bias is extreme as illustrated by $k = 1,2$. This illustration does not consider the issue $SRS_U - GREG$ has with invertible matrices for small samples in combination with many auxiliary variables. In the simulation conducted in section 5.1, it was clear that $SRS_U - GREG$ struggled with the sample sizes yielded in this simulation, therefore the practically achievable sample size needed for $SRS_U - GREG$ is probably around 400 rather than 269 to ensure that the CV as well as a possibility to produce an estimate. We observe that $SRS_B - MADI_{RF}$ is better than the $SRS_U - HT$ regardless of selection bias mechanism and size in $A_{kl}$ for this specific illustration, except for $A_{11}$. This shows that even in the presence of a relatively small NPD with an extraordinary high selection bias, $SRS_B - MADI_{RF}$ is at the very least equivalent to $SRS_U - HT$.

|   | 10% | 20% | 30% | 40% | 50% | 60% | 70% | 80% | 90% |
|---|---|---|---|---|---|---|---|---|---|
| $SRS_U - HT$ | 3044 | 3044 | 3044 | 3044 | 3044 | 3044 | 3044 | 3044 | 3044 |
| $SRS_U - GREG$ | 269 | 269 | 269 | 269 | 269 | 269 | 269 | 269 | 269 |
| $A_{1l} - MADI_{RF}$ | 3081 | 3030 | 2820 | 2850 | 2857 | 2627 | 2386 | 1788 | 984 |
| $A_{2l} - MADI_{RF}$ | 2210 | 1646 | 1266 | 988 | 775 | 626 | 518 | 417 | 255 |
| $A_{3l} - MADI_{RF}$ | 122 | 106 | 97 | 91 | 102 | 112 | 113 | 137 | 139 |
| $A_{4l} - MADI_{RF}$ | 125 | 117 | 97 | 83 | 79 | 73 | 64 | 65 | 42 |
| $A_{5l} - MADI_{RF}$ | 123 | 109 | 80 | 82 | 76 | 65 | 70 | 73 | 76 |
| $A_{6l} - MADI_{RF}$ | 151 | 142 | 140 | 136 | 157 | 157 | 222 | 286 | 211 |
| $A_{7l} - MADI_{RF}$ | 229 | 163 | 119 | 130 | 107 | 87 | 103 | 90 | 80 |
| $A_{8l} - MADI_{RF}$ | 126 | 96 | 89 | 77 | 73 | 75 | 89 | 108 | 91 |

Table 3. The table show the number of units needed to reach a CV of below 1% for the different sampling strategies named in the first column. The following columns illustrate the proportion of U allocated to B. The rows starting with $A_{1l}, \ldots, A_{8l}$ illustrate different generations of $\theta_{kl}$ as described above used to generate the NPD used in the $SRS_B - MADI_{RF}$.

## 6. Conclusion

In this paper we propose the $MADI$ sampling strategy that can incorporate information from a non-probability dataset. Ranalli (2025) highlight the importance of unbiased and reliable methods when using such data to avoid uncertainties introduced by having data not collected by a survey with associated design weights. We have shown both theoretically and empirically that the $MADI$ sampling strategy is unbiased and that it has an unbiased variance estimator. The simulation showed that the proposed strategy performs well for realistic scenarios compared to standard alternatives. One risk with the $MADI$ sampling strategy is that we potentially get higher variance compared to more traditional estimators. This can occur when the underlying selection bias in $A$ is high. However, the simulations shown in this paper suggest that the estimator is quite robust against relatively high selection bias, even when the size of $A$ is relatively low. The estimator is competitive in the sense that it does not rely on MAR in the NPD which is quite common in other estimators that incorporate such data. The estimator is also easy to implement in statistical production, especially for statisticians with a background in survey theory.

A hot topic at many NSO is the costs attached with large surveys, mainly due to large samples, and the associated response burden. We have shown in both simulation 1 and 2 that the number of sampled units needed in the probability sample were substantially lower with the $MADI$-estimator compared to both the GREG estimator and the HT estimator in most situations. This is achieved while we with a design-based estimator which uses traditional design weights to ensure unbiased estimates. The effect of nonresponse in the probability sample have not been studied in this paper and should be further look at. However, assuming the number of sampled units needed in the probability sample is reduced compared to a traditional survey, NSOs may have the capacity to commit more resources towards acquiring the answers from the sampled units and thereby also reducing the nonresponse percentage of the probability sample.

Many surveys at NSOs are conducted on several study variables at each survey period. It may be that different auxiliary variables are more or less optimal for a GREG model for different study variables in the same survey. In the simulations conducted in this paper, the GREG estimator deteriorated when the number of auxiliary variables was large in proportion to the sample size. There are several ways to address this issue. For example, the model can use a reduces and fixed number of auxiliary variables for all study variables. This is probably the most common way to address the issue but may be less effective estimates. Another approach could be to use different auxiliary variables for all study variables. This would most likely lead to more efficient estimates compare to the first option but would be substantially more time consuming. A third option is to use generalized inverse matrices to estimate the model parameters, for example Moore–Penrose pseudoinverse. It can however severely destabilize the weights because small singular values are inverted to disproportionately large values thereby amplifying noise and leading to highly unreliable estimates. It can also be computationally costly to calculate the singular value decomposition if the number of auxiliary variables is substantial. This does not pose a limitation for the $MADI$-estimator if the non-probability dataset is of a sufficient size. There may be situations where the NPD include information about some but not all study variables. Under such conditions, the $MADI$-estimator is still applicable of the study variables where the NPD information is available and the remaining study variables could be investigated with a traditional survey.

The $MADI$ sampling strategy relies on NPD that include information about both the study variable and the auxiliary variable. For many NSOs, this kind of data is available for different surveys, especially for business statistics where there may be situations where data are gathered continuously from certain companies via some process not involving survey methods. Webb panels or voluntary samples are other data sources that might be used in combination with the $MADI$ sampling strategy. Depending on the NPD source, the data may not measure exactly the same variable as the study variable. It is not investigated in this paper but if the information about the study variable in $A$ is deemed unreliable, it would be possible to also draw a probability sample from $A$ and use the same method as described in this paper to adjust for measurement errors. This is addressed in Kim and Tam (2021) should be an avenue for further research for the $MADI$ sampling strategy.

We have limited ourselves to illustrate that the $MADI$-estimator works under certain conditions in this paper. However, further research should look at how the estimator performs with the implementations of other models. A clear advantage of the $MADI$-estimator is that if the dataset $A$ is relatively large, the number of available models becomes larger compared to a traditional survey dataset. For example, a neural network of several dimensions requires a vast amount of data to accurately estimate all model parameters, this is simply not feasible with the amount of data gathered from most surveys. However, the DI-estimator can easily utilize such a model if the size of $A$ is large enough.